# A systematic framework to discover pattern for web spam classification

Hamed Jelodar[1], Yongli Wang[1], Chi Yuan[1], Xiaohui Jiang[2]
Department of Computer Science and Engineering
Nanjing University of Science and Technology,
Nanjing - 210094, China
{Jelodar, Yongliwang, Yuanchi}@njust.edu.cn[1], Jsujiang@jsu.edu.cn[2]

*Abstract*

**Web spam is a big problem for search engine users in World Wide Web. They use deceptive techniques to achieve high rankings. Although many researchers have presented the different approach for classification and web spam detection still it is an open issue in computer science. Analyzing and evaluating these websites can be an effective step for discovering and categorizing the features of these websites. There are several methods and algorithms for detecting those websites, such as decision tree algorithm. In this paper, we present a systematic framework based on CHAID algorithm and a modified string matching algorithm (KMP) for extract features and analysis of these websites. We evaluated our model and other methods with a dataset of Alexa Top 500 Global Sites and Bing search engine results in 500 queries.**

## I. Introduction

The Internet is a cheap and practical tool for humans in various fields that including commerce, entertainment, education, and other. In this Internet community, spam is one of the dangers that threaten users and search engines on the Internet, the web spam or spamdexing concept was first introduced in 1996 [1], and it has been considered as a key challenge for search engines as it affects the quality of the search results. Web spam basically aims at misleading the search engine results.

Nowadays, the number of Internet users has increased, and spammers consider it as a good opportunity to for more revenue. These websites, create a waste of time and cost for Internet users. For example, a user was searching for a useful article on a trusted website by a search engine. However, most of the links in the results are not related to the user's request and redirected to a spam page. These Web sites use various techniques to obtain higher rankings in a search engine's results and attractive of users. There are different approaches and methods to detect spam web that including link-based, content-based and click-based. In general, our approach is based on content and links. Many methods presented in previous works.

Caverlee et al. [2] analyzed the links based on its credibility scores. They defined "k-Scoped Credibility" concept for each page. They also offered several different methods to estimate it and showed how their approach can be effective in detecting web spam. In particular, they first defined the concept of "Bad Path," which accidentally began to work as a "k-hop" from the current page and end up on the spam page. Egele and et al. [3] they proposed an approach to web spam discovery in the list of results that are obtained from a search engine. The authors had chosen a decision tree for classifier for test and evaluation of their data set, so they generated a decision tree by the J48 algorithm that achieved 35% detection rate and 11% false positives. Geraci et al. [4] presented a new approach based on clustering algorithm (M-FPF) to spam web discovery. For evolution their approach, they built a manual data set from these websites (spam) that include 917, 581 non-spam pages and 158, 498 spam pages. They compared their approach with 'LASH-fingerprint clustering algorithm described' and demonstrated that their approach can obtain 0.9640 % in F-measure, and also 0.6925 % for LASH.

Our study focuses on web spam classification and analysis characters on these websites (such as internal links, keywords). This paper provides a framework based on decision tree algorithm for extracting features and patterns of these websites. This framework consists of three steps: First, we considered websites with different issues for inputs that were selected from "Alexa-Top-Websites" as well as using "Bing search engine results" considering 500 questions; then inputs were labeled. The website diagnosis was done by using a modified Knuth–Morris–Pratt(KMP) algorithm in order to find the key parameters. In the next step, the desired features of any website were extracted and saved in a data set after identifying the characteristics of the previous step. Finally, the decision tree type classification was used in order to discover patterns and produce rules after the process of analysis and data preparation; we have considered the CHAID algorithm. This algorithm is a very useful method for identifying the important relationship between independent variables and a dependent variable.

The main contributions of this article are summarized as follows:

- We designed an algorithm for extracting attributes from each website and then obtained the attributes score.

- We analyzed 4272 websites for web spam classifications from Bing search engine results in 500 queries

- We evaluated the performance of the CHAID algorithm and several other machine learning algorithms for annoying web classification.

## II. BACKGROUND

Many researchers presented different approaches for web spam detection and we refer some impressive work. Shen et al. [5] introduced an approach link-based to spam detection with using Temporal information specifically, The authors provided a spam classification model based on SVM algorithm with considering In-link Growth Rate (IGR) and In-link Death Rate (IDR). Hu et al. [6] presented an Intelligent Hybrid Spam-Filtering Framework (IHSFF) to spam detection based on analyzing email headers. The authors extracted five attributes from the email header that including: ''destination field'', "originator field", ''sender server IP address'', ''X-Mailer field" and ''mail subject". They used various machine-learning algorithms to evaluate their model and proved that the Random Forest algorithm achieves the best results in accuracy, precision, recall, and F-measure. Shengen et al. [7], proposed, generating new features with using SVM algorithm and genetic programming for link spam detection. They performed their experiments on WEBSPAM-UK2006 collection.

**Fdez-Glez and et al.** [8], they designed a novel framework for web spams filtering that called WSF2 and used multiple classification algorithms in this framework. They demonstrated that WSF2 is an efficient method to combat with the concept drift problem in spam classification and better than other approaches like SVM and Bagging and Adaboost. They applied HostRank [9] as a ranking mechanism in this approach:

$$HR = \alpha \cdot \sum_{(q,p)\in\varepsilon}\left(\frac{HR(q)}{\deg^+(q)}\right) + (1-\alpha) \cdot \frac{1}{N} \quad (1)$$

Where *HR(p)* is the HostRank result on host *p*, $\alpha$ is a decay factor. Top rank results are then evaluated as spam seed set and selected seeds as spam to form spam vector *B*, where

$$B(p) = \begin{cases} 1 & p \in \upsilon_s \\ 0 & otherwise \end{cases} \quad (2)$$

The normalized spam vector *B* is used later in both the original and modified version of Web spam detection algorithms to propagate distrust to detect more spam.

Liu et al. [10] proposed a new algorithm based a Good-Bad Rank(GBR) Algorithm to combat and identify of spam. in this method, they adopted the GoodRank scores of the demotion from spam and used BadRank scores of GBR for the detection of spam. For evaluation of GBR algorithm, they used WEBSPAM-UK2007 and ClueWeb09 dataset. Experimental results show that GBR better than other typical link-based anti-spam methods (such as PageRank[11], HITS[12]). , the GoodRank (BadRank) scores being propagated is penalized by the page's BadRank (GoodRank), i.e., the propagation of a page's trust/distrust is decided by its probability of being trust/distrust. The formulas are:

$$g(p) = \alpha \cdot \sum_{q\in IN(p)} \frac{g(q)}{OUT(q)} \cdot \frac{g(q)}{q(q)+b(q)} + (1-\alpha).d(p), \quad (3)$$

$$b(p) = \alpha \cdot \sum_{q\in out(p)} \frac{b(q)}{IN(q)} \cdot \frac{b(q)}{g(q)+b(q)} + (1-\alpha').d'(p), \quad (4)$$

Where *g*(p) and *b*(p) are the GoodRank score and BadRank score of page p, respectively.

Geraci et al. [3] presented a new approach based on clustering algorithm (M-FPF) to spam web discovery. For evolution their approach, they built a manual data set from these websites (spam) that include 917, 581 non-spam pages and 158, 498 spam pages. They compared their approach from 'LASH-fingerprint clustering algorithm described' and Experiments show that their approach can get 0.9640 % in F-measure, and 0.6925 % for LASH. Goh et al. [4] proposed a novel approach based on distrust seed set propagation algorithm for detecting web spam, called DSP. They compared DSP with other spam detection algorithms that include (Krishna et al. [13] Anti-TrustRank, Wu et al. [14] distrust algorithm and Nie et al. [15] distrust algorithm). The results have shown that DSP can detect17.73 % more spam hosts of the former dataset (WEBSPAM-UK2006) and can detect 8.59 % more spam hosts of WEBSPAM-UK2007 dataset.

Wu and et al. [16] introduced a unified approach to spam message co-detection and spam messages in microblogge websites (such as; Twitter, SinaWeibo ) . This approach used relations of posting messages and users to merge and spam message discovery and social spammer discovery and used an effective optimization algorithm based on ADMM [17]. For evaluation their approach, they randomly selected 200 users and obtained 53,484 messages from SinaWeibo website . Liu and et al. [10] the authors focused on analysis patterns of User visiting in spam pages. They introduced a novel spam detection framework of detecting types of unknown spam. For approach evaluation, they used a dataset that includes over 800 million Chinese Web pages , according to the obtained results , their framework can analyze large-scale Web access logs . Silva et al. [18] presented a classification method based on the minimum description length (MDL) principle, called MDLClass . An important feature of this approach is the low computational cost with regard to support for complex models. They applied one-hot encoding (OHE), and used a discretization using the well-known entropy based on heuristic proposed by Fayyad and Irani [18], that consider it as:

$$L(d|c_j) = \sum L(t_i|c_j) \times K(t_i), \quad (5)$$

Where,

$L(t_i|c_j) = i - \log^2 \beta(t_i|c_j)$, the term $\beta(t_i|c_j)$ corresponds to the conditional probability of feature $t_i$ given the class $c_j$.

For test and evaluation, they used two standard datasets (real) that include WEBSPAM-UK2007 and WEBSPAMUK2006. The results showed that the proposed approach (MDLClass) can be effective for web spam filtering through considering various types of features.

Goh and et al. [19] suggested a novel method using weight properties to enhance of detection rate for Web spam detection algorithms, and defined the weight properties as an effective item of one Web node to another Web node. Goh and et al. [20] proposed a machine learning approach based on multilayer perceptions (MLP) neural network Algorithm to improve the accuracy of Web spam detection. They compared their approach with support vector machine (SVM) and they have shown that MLP networks better than SVM up to 14.02% on WEBSPAM-UK2006 dataset and up to 3.53% on WEBSPAM-UK2007 dataset.

Egele and et al. [3] proposed an approach to web spam discovery in the list of results that is obtained from a search engine. The authors had chosen a decision tree for classifier for test and evaluation of their data set. So they generated a decision tree by a J48 algorithm that achieved 35% detection rate and 11% false positives. Ahmad et al. [21] suggested a generic statistical approach based on Spam profile detection on Facebook and Twitter networks and utilized three various classification algorithms that include naive Bayes, Jrip, and J48 for analyzing discriminative properties of features. They demonstrated that J48 algorithm is the best classification for both Facebook and Twitter dataset.

### III. PROPOSED APPROACH

The classification rules of the decision tree structure are deducted from irregular and random samples. Basically, decision trees are developed to provide a tree-based approach on a series of statistical data for the classification of similar statistical data. Decision trees are composed of a simple structure. The non-terminal nodes include one or more attributes, and the terminal nodes reflect the output results. A decision tree was designed from right to left, and the output was developed from the root node towards the bottom. Decision trees provide a remarkable prediction and a conceptual description of the data set [22, 23]. There are many decision tree algorithms, including CART, C4.5, and CHAID. In this study, the CHAID algorithm was used for classification purposes. However, this process is not simple. In the first stage, using the causal relationship exploration, the level of independent variables is classified for a given dependent variable. Moreover, the CHAID algorithm provides interaction between many independent variables.

According to this algorithm, a framework is provided for feature extraction and analysis of spam features as shown in Figure 1. An algorithm was also provided to extract features from any website and obtain the points of each parameter. This algorithm has been shown in Table 1. First, a set of templates were considered as input and the KMP algorithms and Regular Expressions technique were used to achieve them. This technique can have an effective role in the identification of data content parameters. For example, this technique is efficient in obtaining the parameter of "keywords" tag on the web. According to this algorithm, a code line ranging from 3 to 5 means that the data contents are taken from a Web page and stored in *s_data*. A code line ranging from 6 to 9 means that the search and calculation of points for each feature are under process.

TABLE 1. FIND AND GET SCORE FOR KEYWORDS BASE OF KNUTH–MORRIS–PRATT ALGORITHM

| Algorithm 1: Find and get score for keywords base of KMP algorithm | |
|---|---|
| Input: | A set of pattern variables X={p1,p2,...,pn}, |
| Output: | An integer value that returns score parameters |
| | **Begin** |
| 1: | scr_p= a variable for save of score |
| 2: | **for** j_1 **to** row_lurl.count // there are list of url in row_lurl |
| 3: | s_data = get_HtmlData(j_1) |
| 4: | s_data =s_data.gbody() // return of content from body html |
| 5: | s_t[j_1] =s_data //' s_t' is a varable |
| 6: | **If** sk_kmp (s_t[i],k_w[i]))== true **then** |
| 7: | scr_p + = 10; // we consider a score for value of keywords(spical key and public), for example : value for sepical key_word =10; |
| 8: | **else if** Gscr_regEx(ht[i] ,p[i])== true **then** |
| 9: | scr_p + = 10; |
| 10: | **End if** |
| 11: | **End for** |
| 12: | **End** |

Time complexity is one of the most important criteria for detection of algorithms efficiency. The purpose of calculating the time complexity of an algorithm was to quantify the amount of time taken by the algorithm to run. In order to calculate time complexity of algorithms, the number of algorithm steps was identified as a function of the problem size. To this end, the frequency of the original algorithm's operation is calculated and expressed as a function f. After that, function g, which indicates the magnitude order of the function f when the input size is large enough, was obtained. Therefore, criteria such as time complexity are required to detect the efficiency of algorithms. The analyzed time complexity for this algorithm is in the *O (N)* order.

*This framework consists of three phases, which include:*

**First Phase:**

Websites on different topics are considered as input. These websites are selected and derived from the "Top 500 websites Alexa" as well as "Bing search engine results" based on 500 query in different subject. The inputs are then compared to the blacklist (labeling on the inputs). Key words, which are usually embedded on web pages using different techniques, serve as one of the effective parameters for identifying the nature of spam websites. For example, some websites usually use specific and tricky key words. In this case, the KMP algorithm is a string algorithm that can be used for pattern matching and obtain value of attributes for prepare our dataset [27]. When a shift takes place in a text or two patterns are matched, this algorithm uses the data obtained from a special table to avoid the characters that have already been checked and therefore, limits the number of comparisons. Based on the above information, it is useful to identify the key words and use this algorithm to match and identify them.

**Second Phase:**

Feature extraction is a very commonly used process in data processing. In this step, some operations are carried out on the data to identify their decisive and distinctive features. During this process, the raw data take more usable forms for the subsequent statistical processes. In fact, the data noises are isolated, and the desired features are extracted. Algorithm 1 plays an important role in the successful conduction of this process.

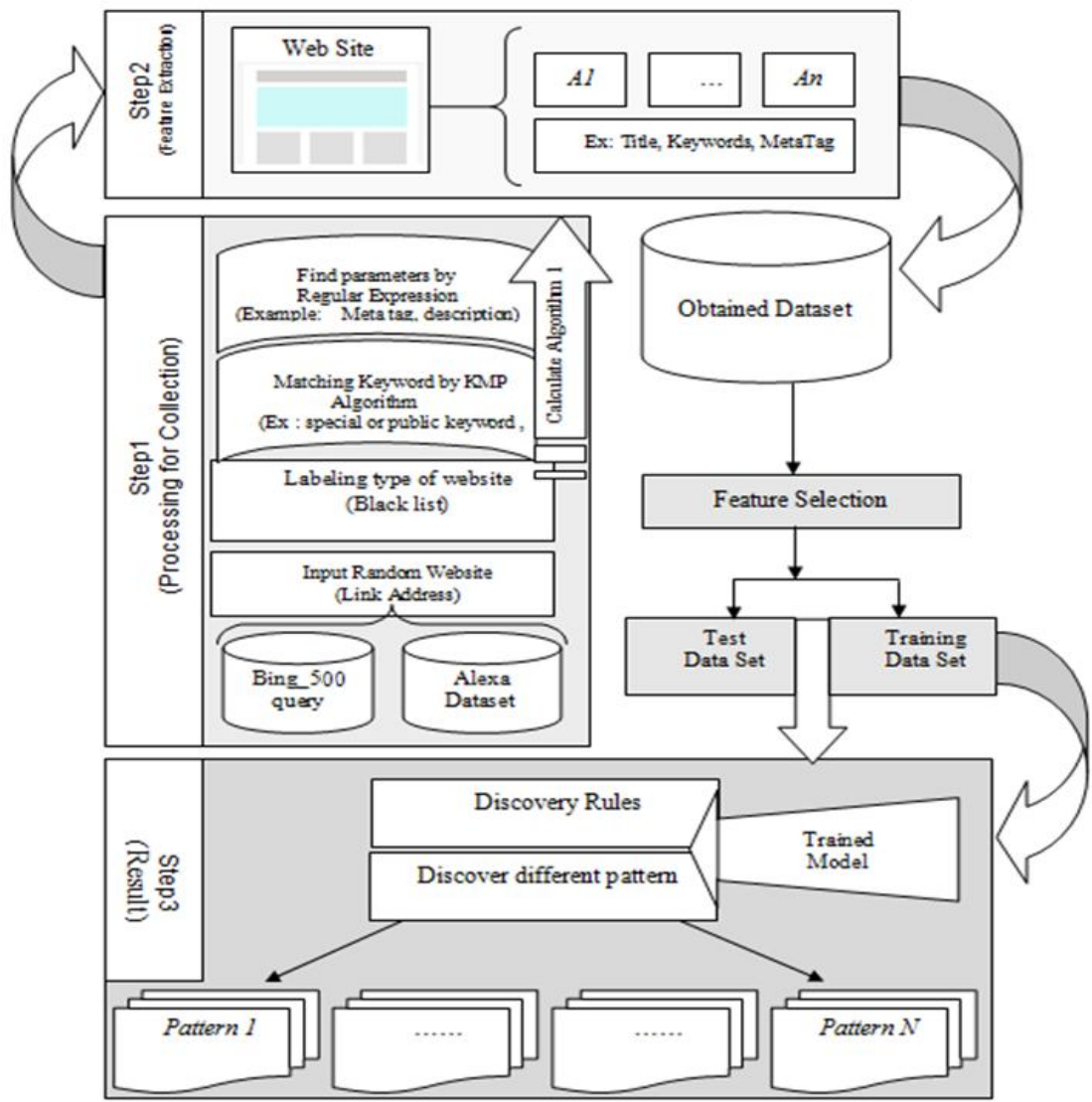

Fig. 1. Proposed research model based on decision tree methodology

**Third Phase:**

After the process of data analysis and preparation, a decision-tree classification is used to discover patterns and generate rules. The CHAID algorithm is considered for this purpose. This algorithm is useful in identifying the relationship between the independent and dependent variables. In this step, the analysis of these variables results in the development of a predictive model, and using the model, different behavioral patterns can be obtained between the nodes. For example, in our test, we found that the websites with a score of (keywords of special= max and keyword of public= very max) can probably be regarded as a spam. As previously mentioned, the CHAID algorithm can have a major influence on the factors that accelerate the extraction process. In CHAID trees, the homogeneity of the groups generated by the tree is evaluated by a Bonferroni corrected p-value obtained from the chi-square statistical test applied to two-way classification tables with C classes and K splits for each tree node as presented below [25]:

$$X^2 = \sum_{j=1}^{J}\sum_{i=1}^{I} \frac{(n_{ij} - m_{ij})^2}{m_{ij}} \qquad (6)$$

*where,*

$$n_{ij} = \sum_{n \in D} fnl(x_n = i \cap y_n = j) \qquad (7)$$

The performance procedure of this algorithm is as follows:

IV. RESULTS AND STATICAL ANALYSIS

The statistical analysis was performed by using SPSS version 19 (IBM Corp., USA) in order to evaluate and design statistical data sets (it provides different classifications of algorithms such as: CHAID, C5). This tool is responsible for automatic splitting of training and test data. The aim of splitting is to produce a model quality test mechanism [24, 26].

*Step 1:* Assuming that Y is the target variable, for each explanatory variable X, two categories with the least difference in terms of distribution can be found. The methods used for the amount of P depend on the type of target variable Y. If the target variable Y is continuous, the F test is used. If the target variable Y is nominal, a D-2 cross-classification table will be formed by columns X and row Y, and the Pierson χ2 (Pearson chi--square) and likelihood ratio test are used. If the target variable Y is sequential or discrete, the likelihood ratio test is used.

*Step 2:* Two sets of X are found through maximum P value and then compared to the amount of pre-set merge P at the α-merge level.

If P value is less than α- merge, then the two sets of X are merged and a new set of X series will be formed. Then step 1 is repeated. If P value is larger than α-split, we need to go for the third step.

*Step 3:* The Bonferroni method is used to calculate the amount of P set to the likelihood table by the explanatory variable X and the target variable Y.

*Step 4:* The explanatory variable X is selected by setting the least amount of P, and the P value is compared to the pre-set level of α- split.

If the value of P is less than the α-split, then the node is split by this Category of X series; if the value of P is greater than the α-split, then the node will not be split and will only be a terminal node.

*Step 5:* Decision tree continues to grow until it faces the stop rules.

Models were created using training data, and the test data was used to estimate the quality. We considered splitting nodes: 0.05, Merging Categories: 0.05, training = 70% and test data = 30 to test this study. Table 2 shows the attributes of the considered in the dataset. The value range based on the experimental test has been obtained.

TABLE 2. ATTRIBUTES THAT HAVE CONSIDERED FOR THE DATASET

| No. | Attribute in dataset | Description |
|---|---|---|
| 1 | Black list | Check links in a black list |
| 2 | Feature of URL | This attribute check with a series of selected key words. |
| 3 | Meta_tak | This attribute check with a series of selected key words. |
| 4 | Key_word_special | Check special keywords in first page a website. Ex: adult words , advertising words |
| 5 | Key_word_public | Check public keywords in first page a website. |
| 6 | Count of_internal_link | Check the count of internal websites |
| 7 | Count _external_link | Check the count of external links to websites |
| 8 | Count of_post | Check the count of page post to websites |

We used two datasets with real data to test this study. The datasets were prepared manually due to some limitations.

**Bing Search Result:** For consideration of various issues, we used 500 queries. The queries were entered manually into "Bing search engine." The search results were between 20 and 70 links. Total collected data included 4272 Web spam and non-spam websites.

**Alexa_top_web:** Alexa is one of the most famous companies that are active in website analysis and ranking, and it can be used as a standard reference. In fact, its main task is ranking websites based on their consumption traffic. It determines the web rank depending on the amount of consumed bandwidth. It includes a collection of websites with different themes, so we referred to Alexa.com site and randomly selected the Web sites as our input.

In general, all experiments have been run on a machine with Core i2 CPU and 3G of RAM. To test and evaluate the algorithms, we used k-fold cross validation. In this process, the data set was divided into k subsets. Each time, one of the k subsets was used as the test set, and the other k-1 subsets form the training set. Performance statistics were calculated across all k trials. Data were tested and analyzed by the use of the framework. We assessed the data set by qualitative criteria, which mainly are Recall, F-measure and Precision.

This framework base of CHAID algorithm was successfully employed in producing results that effectively classified the data into attributes whose values were individually determined. As it can be seen, we have 4272 links including 3199 non-spam and 1073 spam. According to figures 2 and 3, we can find some patterns based on the decision tree generated by the proposed framework:

**Pattern A:** *Classification of web spam based on keywords of special and keywords of public*, CHAID algorithm created a tree model with branches for websites based on keywords of special and keywords of public. The tree model consists of three levels.

*Level 1:* there are multiple nodes at this level. More spams are observed in node 4. According to Node 1, if value of keyword of special = max then we have 826 websites that are rated as "true spam = 78.9%" and "false spam = 21.1%" in first level.

*Level 2:* We have keyword_public attribute at second level. We have four nodes at this level. More spams are observed in node 4. According to Node 4, we can see that if value of keywords of special= max and keyword of public= very max then true spam is 87.1%. we can see that these two attribute with each other are impressive for spam classification.

**Pattern B:** Classification of webspam based on Count of internal link and Count of external link, the second tree model indicates a branching for detection of spam based on Count of internal link and Count of external link. The tree model consists of three levels.

*Level 1:* According to Node 1 we have, if value of "count of internal link" = max, we have 1431 websites that are rated as "true spam = 67%" and "false spam = 33%".

*Level 2:* According to Node 5, we have 1071 websites that if value of count of internal link is max and count of external links is max Then true spam is 75.9% and false spam is 24.1%. As before, we can see that these two attribute with each other are impressive for spam classification.

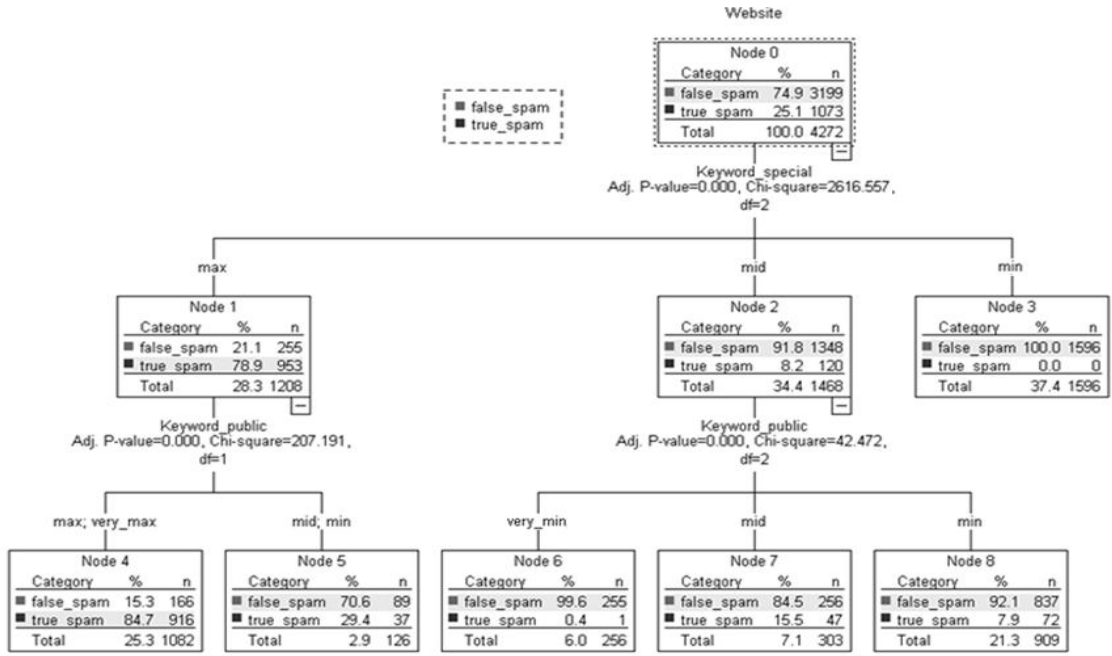

Fig. 2. Classification of web spam based on keywords of special and keywords of public

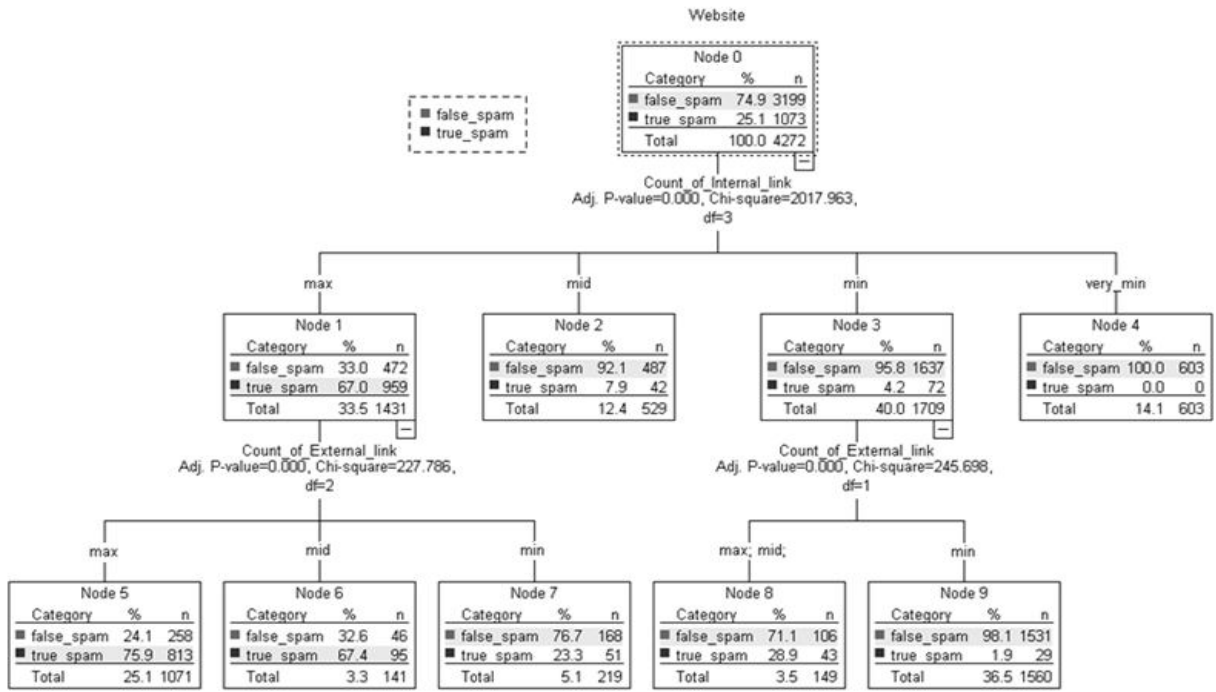

Fig. 3. Classification of web spam based on count of internal links and count of external links

## V. CONCLUSION

In this paper, we presented a systematic framework based on CHAID algorithm and a modified string matching algorithm (KMP) for extract features and analysis of these websites. For our experiments, we considered websites with the different topic for inputs that were selected from Alexa Top 500 Global Sites and Bing search engine results in 500 queries, and according to our approach, we generated a tree model with various attributes that can discover some of the influential patterns of web spam classification. For example, in Patten B (fig 3), we showed the attributes of 'count of internal link ' and 'count of external link ' with each other are impressive to web spam classification. Definitely, the detection of web spam and the separation and deletion of such webs from the search engine results can have a high effect on the optimization of search results and also prevents the Internet users from having unwanted access to these webs. Our main goal for this presented framework is an intelligent analysis to find effective attributes and classification of the websites.

ACKNOWLEDGMENT

The authors would like to thank the anonymous reviewers for their valuable comments and suggestions. This work is supported in part by the National Natural Science Foundation of China under Grant 61170035 and 61272420, Six talent peaks project in Jiangsu Province (Grant No. 2014 WLW-004), the Fundamental Research Funds for the Central Universities (Grant No. 30916011328), Jiangsu Province special funds for transformation of science and technology achievement (Grant No. BA2013047).